\documentclass[a4paper,11pt]{article}

\usepackage[utf8x]{inputenc}
\usepackage[centertags]{amsmath}
\usepackage{amssymb,amsthm,epsfig,psfrag}
\usepackage{graphicx}
\usepackage{dsfont}
\usepackage{color}
\usepackage{pstricks-add}
\usepackage{fullpage}
\usepackage{etex}

\usepackage{bm}
\usepackage{epic}
\usepackage{hyperref}
\usepackage{young}
\usepackage{multirow}
\usepackage{fancybox}

\newcommand{\sfrac}[2]{{\textstyle\frac{#1}{#2}}}

\newcommand{\alg}[1]{\mathfrak{#1}}

\usepackage{fix-cm}

\usepackage[nosort]{cite}
\usepackage[bulletsep]{collref}
\usepackage{tensor}

\begin{document}

\thispagestyle{empty}

\begingroup\raggedleft\footnotesize\ttfamily
HU-EP-14/26\\ 
AEI-2014-030\\
HU-Mathematik-2014-16\\
CERN-PH-TH-2014-107\\
LMU-ASC 42/14

\endgroup

\begin{center}
{\Large\bfseries $\mathcal{N}=4$ Scattering Amplitudes and the Deformed Gra{\ss}mannian\par}%
\vspace{15mm}

\begingroup\scshape\large 
Livia Ferro${}^{1}$, Tomasz \L ukowski${}^{2}$,
Matthias Staudacher${}^{3,4,5}$
\endgroup
\vspace{7mm}

\textit{
${}^1${
Arnold-Sommerfeld-Center for Theoretical Physics\\ Fakult\"at f\"ur Physik, Ludwig-Maximilians-Universit\"at M\"unchen, \\Theresienstra\ss e 37, 80333 M\"unchen, Germany}\\[0.4cm]
${}^2${
Mathematical Institute, University of Oxford, Andrew Wiles Building,\\ Radcliffe Observatory Quarter, Oxford, OX2 6GG, United Kingdom}\\[0.4cm]
${}^3${
Institut f\"ur Mathematik, Institut f\"ur Physik und IRIS Adlershof,\\
Humboldt-Universit\"at zu Berlin,\\
Zum Gro\ss en Windkanal 6, 12489 Berlin, Germany}\\[0.2cm]
${}^4${
Max-Planck Institut f\"ur Gravitationsphysik, Albert-Einstein-Institut,\\
Am M\"uhlenberg 1, 14476 Potsdam, Germany}\\[0.2cm]
${}^5${
Theory Group, Physics Department, CERN,\\
1211 Geneva 23, Switzerland}
}\\[0.4cm]

{\tt staudacher$\bullet$mathematik.hu-berlin.de,\\
lukowski$\bullet$maths.ox.ac.uk\\
livia.ferro$\bullet$lmu.de}\\[0.2cm]

\vspace{8mm}

\textbf{Abstract}\vspace{5mm}\par
\begin{minipage}{14.7cm}

Some time ago the general tree-level scattering amplitudes of ${\cal N}=4$ Super Yang-Mills theory were expressed as certain Gra{\ss}mannian contour integrals. These remarkable formulas allow to clearly expose the super-conformal, dual super-conformal, and Yangian symmetries of the amplitudes. Using ideas from integrability it was recently shown that the building blocks of the amplitudes permit a natural multi-parameter deformation. However, this approach had been criticized by the observation that it seemed impossible to reassemble the building blocks into Yangian-invariant deformed non-MHV amplitudes. In this note we demonstrate that the deformations may be succinctly summarized by a simple modification of the measure of the Gra{\ss}mannian integrals, leading to a Yangian-invariant deformation of the general tree-level amplitudes. 
Interestingly, the deformed building-blocks appear as residues of poles in the spectral parameter planes.
Given that the contour integrals also contain information on the amplitudes at loop-level, we expect the deformations to be useful there as well. In particular, applying meromorphicity arguments, they may be expected to regulate all notorious infrared divergences. We also point out relations to Gelfand hypergeometric functions and the quantum Knizhnik-Zamolodchikov equations.

\end{minipage}\par
\end{center}
\newpage

\section{Motivation and Results}

Nature prefers Yang-Mills theory in exactly $1+3$ dimensions. There has been much recent interest in a mathematically exceedingly rich four-dimensional Yang-Mills model, the nearly unique ${\cal N}=4$ supersymmetric theory \cite{Brink:1976bc,Gliozzi:1976qd}. In addition to its gauge and super-conformal symmetries, it exhibits, in the planar limit, the phenomenon of {\it integrability}, see the series of review papers \cite{Beisert:2010jr}. What is special about $1+3$ dimensions? One remarkable fact is that general space-time events with Minkowski coordinates $x^\mu \in \mathbb{R}^{1,3}$ may be packaged into general $2 \times 2$ hermitian matrices. After Fourier-transforming to dual space-time, a momentum four-vector $p^\mu \in \mathbb{R}^{1,3}$ may be written as the hermitian matrix
\begin{equation}
p^{\alpha \dot \alpha} =
\begin{pmatrix}
p_0+\, p_3& p_1-i\, p_2\\
p_1+i\, p_2&p_0-\, p_3
\end{pmatrix}.
\end{equation}
Massless particles satisfy $p^2= p^\mu p_\mu=\det p^{\alpha \dot \alpha}=0$. The matrix then has at most rank $1$, and we can ``factor'' it into spinorial Weyl variables: $p^{\alpha \dot \alpha}=\lambda^\alpha \tilde \lambda^{\dot \alpha}$. For ${\cal N}=4$ super Yang-Mills the spinors $\lambda^\alpha$, $\tilde \lambda^{\dot \alpha}$ are nicely complemented by the four Gra{\ss}mann spinor variables $\eta^A$ with $A=1,2,3,4$. The resulting eight spinor-helicity variables $(\lambda^\alpha_j, \tilde \lambda^{\dot \alpha}_j, \eta^A_j)$ are highly efficient for neatly expressing the general color-stripped tree-level amplitudes for the scattering of $j=1, \ldots, n$ massless particles of the model. With total momentum $P^{\alpha \dot \alpha}=\sum_j \lambda^{\alpha}_j \tilde \lambda^{\dot \alpha}_j$ and super-momentum $Q^{\alpha A}=\sum_j \lambda^{\alpha}_j \eta_j^A$ and the brackets 
$\langle p q \rangle=\epsilon_{\alpha \beta} \lambda_p^\alpha \lambda_q^\beta$
and
$ [p q]=\epsilon_{\dot \alpha \dot \beta} \tilde \lambda_p^{\dot \alpha} \tilde \lambda_q^{\dot \beta}$
the result is the distribution
\begin{equation}\label{general-tree}
{\mathcal A}_{n,k} =
\frac{\delta^4(P^{\alpha \dot \alpha}) \delta^8(Q^{\alpha A})}{\langle 1 2\rangle \langle 2 3\rangle \ldots \langle n-1, n\rangle \langle n 1\rangle}\,{\cal P}_n(\{\lambda_j,\tilde \lambda_j,\eta_j\}),
\end{equation}
see \cite{Drummond:2008cr} and references therein. All external helicity configurations are generated by expansion in the $\eta_j^A.$ Super-helicity $k$ corresponds to the terms of order $\eta^{4 k}$. In the simplest, maximally-helicity-violating (MHV) case we have $k=2$, where ${\cal P}_n=1$. We may also define ``nextness'' $\hat k=k\!-\!2$. Then, for $k>2$, ${\mathcal A}_{n,k}$ corresponds to the N$^{\hat k}$MHV (pronounced ``Next-to-the-$\hat k$-MHV'') amplitude, where the ${\cal P}_n$ are recursively determined rational functions of the spinor helicity variables.

In \cite{ArkaniHamed:2009dn} a remarkable reformulation of \eqref{general-tree} was presented. It takes the form of an integral over a Gra{\ss}mannian space ${\rm Gr}(k,n)$. The latter is the set of $k$-planes intersecting the origin of an $n$-dimensional vector space. Note that $k=1$ is ordinary projective space. ``Points'' in ${\rm Gr}(k,n)$ are described by ``homogeneous'' coordinates, which are packaged into a $k \times n$ matrix $C=(C_{aj})$. Here $C$ and $A \cdot C$ with $A \in {\rm GL}(k)$ correspond to the same point in ${\rm Gr}(k,n)$. It is convenient to employ super-twistors $\mathcal{W}_j^{\mathcal{A}}=(\tilde \mu^{\alpha}_{j},{\tilde \lambda}^{\dot\alpha}_{j}, \eta^{A}_{j})$, where ${\mathcal A}=(\alpha, \dot \alpha, A)$ 
and $\alpha, \dot \alpha=1,\ldots,2$,  $A=1,\ldots,4$,
by performing a formal half-Fourier transform from $\lambda_{j}^{\alpha}$ to $\tilde \mu^{\alpha}_{j}$. The Gra{\ss}mannian integral then reads
\begin{equation}\label{ACCK}
{\mathcal A}_{n,k} =
\int \frac{ d^{k\cdot n} C}{{\rm vol}({\rm GL}(k))}\,
\frac{\delta^{4k|4k}(C\cdot\mathcal{W})}{(1,\,...\,, k)(2,\,...\,, k\!+\!1)\ldots(n,\,...\,, n\!+\!k\!-\!1)}\, .
\end{equation}
The $(i, i+1,\ldots,i\!+\!k\!-\!1)$ are the $n$ cyclic $k \times k$ minors of the coordinate matrix $C$. 
Note that $(n,\,...\,, n\!+\!k\!-\!1) = (n,\,...\,, k\!-\!1)$.
Integration is along ``suitable contours''. The ${\rm GL}(k)$ symmetry is manifest.
Fourier-transforming back to spinor-helicity space, all tree-level N$^{(k-2)}$MHV amplitudes may then indeed be obtained if the contours are correctly chosen. The amplitudes ${\mathcal A}_{n,k}$ enjoy superconformal symmetry
\begin{equation}\label{psu-symmetry}
J^{\mathcal{A} \mathcal{B}} \cdot {\mathcal A}_{n,k}=0\, , \quad {\rm with} \quad J^{\mathcal{A} \mathcal{B}}\in\alg{psu}(2,2|4).
\end{equation}
However, there is also a hidden dual super-conformal symmetry of the tree-level amplitudes
\begin{equation}\label{dual-psu-symmetry}
\tilde J^{\mathcal{A} \mathcal{B}} \cdot {\mathcal A}_{n,k}=0\, , \quad {\rm with} \quad \tilde J^{\mathcal{A} \mathcal{B}}\in\alg{psu}(2,2|4)^{\rm{dual}}\,.
\end{equation}
Commuting $J$ and $\tilde J$, one obtains Yangian symmetry \cite{Drummond:2009fd}. The latter is generated by an infinite algebra consisting of the level-zero generators $J^{\mathcal{A} \mathcal{B}}$ and a set of level-one generators $\hat J^{\mathcal{A} \mathcal{B}}$, plus an infinite tower of further symmetry generators of higher levels, which satisfy certain Serre relations. Using $\alg{psu}(2,2|4)$ generators in super-twistor form acting ``locally'' on the $j$-th particle 
\begin{equation}\label{psu-generators}
J^{\mathcal{A} \mathcal{B}}_j=\mathcal{W}_j^{\mathcal{A}} \frac{\partial}{\partial \mathcal{W}_j^{\mathcal{B}}} - \sfrac{1}{8} (-1)^{\mathcal B} \delta^{\mathcal{A} \mathcal{B}} \sum_{\mathcal{C}} (-1)^{\mathcal C} \mathcal{W}_j^{\mathcal{C}} \frac{\partial}{\partial \mathcal{W}_j^{\mathcal{C}}}\,,
\end{equation}
where the second term removes the supertrace from $\alg{psu}(2,2|4)$ (this is related to the letter $\alg{s}$ in $\alg{psu}(2,2|4)$),
one may succinctly summarize the Yangian algebra relevant to amplitudes as
\begin{equation}\label{yangian-symmetry}
J^{\mathcal{A} \mathcal{B}} = \sum_{j=1}^n J^{\mathcal{A} \mathcal{B}}_j
\quad {\rm and} \quad
\hat J^{\mathcal{A} \mathcal{B}} = \sfrac{1}{2} \sum_{i<j} (-1)^{\mathcal C} \left[J^{\mathcal{A} \mathcal{C}}_i J^{\mathcal{C} \mathcal{B}}_j
- J^{\mathcal{A} \mathcal{C}}_j J^{\mathcal{C} \mathcal{B}}_i\right].
\end{equation}
This is how integrability first appeared in the planar scattering problem. To exhibit the hidden dual symmetry \eqref{dual-psu-symmetry} of the Gra{\ss}mannian integral \eqref{ACCK}, a clever change of variables was found in \cite{ArkaniHamed:2009vw}. Employing $4|4$ super momentum-twistors 
$\mathcal Z_j^{\mathcal A}=(Z_j^\alpha,\chi_j^A)$ with ${\cal A}=(\alpha,A)$ and $\alpha=1,\ldots,4$,  
$A=1,\ldots,4$, one transforms \eqref{ACCK} to an integral over the points of a dual Gra{\ss}mannian space ${\rm Gr}(\hat k,n)={\rm Gr}(k\!-\!2,n)$
\begin{equation}\label{MS}
{\mathcal A}_{n,k} =
\frac{\delta^4(P^{\alpha \dot \alpha}) \delta^8(Q^{\alpha A})}{\langle 1 2\rangle \langle 2 3\rangle \ldots  \langle n 1\rangle}
\int \frac{ d^{\hat k\cdot n} \hat C}{{\rm vol}({\rm GL}(\hat k))}\,
\frac{\delta^{4 \hat k|4 \hat k}(\hat C\cdot\mathcal{Z})}{(1,\,...\,, \hat k) \ldots(n,\,...\,, n\!+\! \hat k\!-\!1)}\, ,
\end{equation}
where the $k=2$ MHV part neatly factors out. 
One has $(n,\,...\,, n\!+\! \hat k\!-\!1)=(n,\,...\,, \hat k\!-\!1)$.
This Gra{\ss}mannian integral based on dual momentum-twistors had been independently discovered in \cite{Mason:2009qx}. Clearly it computes the function ${\cal P}_n(\{\lambda_j,\tilde \lambda_j,\eta_j\})$ in \eqref{general-tree}.

Much of the above beautiful structure is intimately tied to four dimensions. At loop level, infrared divergences appear. These are commonly dealt with by dimensional regularization. However, deviation from four dimensions irretrievably destroys all of the above structure. One is then led to look for a more natural regulator, where natural means it should a) respect the fixed space-time dimensionality four and b) respect the Yangian symmetry, i.e.\ integrability. Such a regularization scheme was proposed in \cite{Ferro:2012xw,Ferro:2013dga}. It may be understood as follows. We should look at the ordinary (as opposed to super) trace in \eqref{psu-generators}. Define the ``local'' and ``overall'' central charge operators,
the minus sign being a convention, respectively as 
\begin{equation}\label{centralcharge}
C_j=-\sum_{\mathcal A} J^{\mathcal{A} \mathcal{A}}_j=
-\sum_{\mathcal A} \mathcal{W}_j^{\mathcal{A}} \frac{\partial}{\partial \mathcal{W}_j^{\mathcal{A}}},
\qquad \qquad
C=\sum_{j=1}^n C_j
 \,.
\end{equation}
These are related to the letter $\alg{p}$ in $\alg{psu}(2,2|4)$. We should ``locally'' or ``overally'' impose $C_j=0$ and $C=0$, respectively, to obtain local or overall $\alg{psu}(2,2|4)$ symmetry, and not some central extension of it. The idea in \cite{Ferro:2012xw,Ferro:2013dga} was to do away with local invariance, and to just impose the overall one. This maneuver has an interesting mathematical as well as physical interpretation. {\it Mathematically}, we are led to the so-called evaluation representation of the Yangian, where \eqref{yangian-symmetry} is modified to
\begin{equation}\label{deformed-yangian-symmetry}
J^{\mathcal{A} \mathcal{B}} = \sum_{j=1}^n J^{\mathcal{A} \mathcal{B}}_j
\quad {\rm and} \quad
\hat J^{\mathcal{A} \mathcal{B}} = \sfrac{1}{2} \sum_{i<j} (-1)^{\mathcal C} \left[J^{\mathcal{A} \mathcal{C}}_i J^{\mathcal{C} \mathcal{B}}_j
- J^{\mathcal{A} \mathcal{C}}_j J^{\mathcal{C} \mathcal{B}}_i\right]-\sum_{j=1}^n v_j J^{\mathcal{A} \mathcal{B}}_j\,.
\end{equation}
``Switching on'' non-zero eigenvalues $c_j$ for the deformed local central charges $C_j$ results in non-vanishing evaluation (or spectral) parameters $v_j$. We will momentarily give the relation between the $c_j$ and the $v_j$, see \eqref{cyclic-identification} below. {\it Physically}, we can interpret the procedure by rewriting the $C_j$ of \eqref{centralcharge} in terms of spinor-helicity variables. One finds
\begin{equation}\label{centralcharge-spinhel}
C_j=2+\lambda^\alpha_j \frac{\partial}{\partial \lambda^\alpha_j}
-\tilde \lambda^{\dot \alpha}_j \frac{\partial}{\partial \tilde \lambda^{\dot \alpha}_j}
-\eta^A_j \frac{\partial}{\partial \eta^A_j}=2-2\,h_j\,
\end{equation}
where $h_j$ is the super-helicity of particle $j$. So we are deforming the helicities of the scattering particles. This is algebraically, read ``locally," consistent, since the quantization of helicities to integer or half-integer values is due to global properties of the conformal group. One could then ask how the Gra{\ss}mannian contour formulas are deformed. The final answer is exceedingly simple, and very natural. Let us define shifted spectral parameters \cite{Beisert:2014qba}
\begin{equation}\label{v-plus-minus}
v_j^\pm=v_j\pm\sfrac{c_j}{2}\,.
\end{equation}
As we will prove in section \ref{sec:Furtherdetails}, one then finds that \eqref{ACCK} is elegantly deformed to
\begin{equation}\label{ACCK-deformed}
{\mathcal A}_{n,k}\left(\{v_j^\pm\}\right)=
\int \frac{ d^{k\cdot n} C}{{\rm vol}({\rm GL}(k))}
\frac{\delta^{4k|4k}(C\cdot\mathcal{W})}{(1,\,...\,, k)^{1+v_k^+-v_1^-}
\ldots(n,\, ...\, ,k\!-\!1)^{1+v_{k-1}^+-v_n^-}}\,.
\end{equation}
Note that it is not really the Gra{\ss}mannian space ${\rm Gr}(k,n)$ as such that is deformed, but the integration measure on this space. One easily sees that the ${\rm GL}(k)$ symmetry of \eqref{ACCK} is also preserved: The measure times delta function factors are ${\rm SL}(k)$ invariant, and so are the minors. Finally, invariance under an overall scale transformation of $C$ is ensured by the telescoping property of the deformation weights on the minors and the vanishing of overall central charge. 
We will show below that formula \eqref{ACCK-deformed} is Yangian invariant, iff we impose $n$ conditions on the $2 n$ deformation parameters $\{v_j^\pm\}$:
\begin{equation}\label{cyclic-identification}
v^+_{j+k}=v^-_j
\qquad {\rm for} \qquad j=1,\ldots,n\,.
\end{equation}
One may then ask, whether the change of variables allowing to go from \eqref{ACCK} to \eqref{MS} still goes through under the deformation. 
Using the following relation from \cite{ArkaniHamed:2009vw} between the minors of the matrices $C$ and $\hat C$
\begin{equation}\label{minor-change}
(i+1,\, ...\, i+\hat k)_{\hat C}
=
\frac{(i,\, ...\, i+k-1)_C}{\langle i,i+1\rangle \ldots \langle i+k-2,i+k-1\rangle}
\,,
\end{equation}
where the subscripts indicate which matrix we consider when evaluating the minors, one easily proves that \eqref{MS} deforms into
\begin{equation}\label{MS-deformed}
{\mathcal A}_{n,k}\left(\{v_j^\pm\}\right)=
\frac{\delta^4(P_{\alpha \dot \alpha}) \delta^8(Q_\alpha^A)}{\langle 1 2\rangle^{1+v_2^+-v_1^-}  \ldots  \langle n 1\rangle^{1+v_1^+-v_{n}^-}}\times A_{n,k}\left(\{v_j^\pm\}\right),
\end{equation}
with
\begin{equation}\label{MS-integral}
A_{n,k}\left(\{v_j^\pm\}\right)= \int \frac{ d^{\hat k\cdot n} \hat C}{{\rm vol}({\rm GL}(\hat k))}\,
\frac{\delta^{4 \hat k|4 \hat k}(\hat C\cdot\mathcal{Z})}{(1,\,...\,, \hat k)^{1+v_{\hat k+1}^+-v_n^-} \ldots(n, \,...\,, \hat k\!-\!1)^{1+v_{\hat k}^+-v_{n-1}^-}}\,.
\end{equation}
Note that both the MHV-prefactor and the contour integral are deformed. From \eqref{cyclic-identification}, we see that the total number of deformation parameters is $k$-independent and equals $n\!-\!1$, since \eqref{ACCK-deformed},\eqref{MS-deformed} depends only on differences of the $\{v_j^\pm\}$.

\section{Meromorphicity Lost and Gained}
\label{meromorphicity}

Let us take a closer look at the deformed Gra{\ss}mannian integrals \eqref{ACCK-deformed} and \eqref{MS-deformed},\eqref{MS-integral}, and compare them to their undeformed versions \eqref{ACCK},\eqref{MS}. The latter have poles in the integration variables $C_{aj}$ or $\hat C_{aj}$, related to the vanishing of the minors. Apart from the delta functions, the integrand is meromorphic, or even better, just a rational function. In contrast, choosing the parameters  $\{v_j^\pm\}$, constrained by \eqref{cyclic-identification}, to be non-integer, we see that generically all poles turn into branch points. Meromorphicity is lost. This does not seem to cause a problem for the MHV amplitudes, where, at least formally, we simply obtain a deformed Parke-Taylor formula, namely the prefactor of the integral in \eqref{MS-deformed}. However, for non-MHV amplitudes with $\hat k >0$, some integrations remain. In the undeformed case, these integrations are performed by the residue theorem. Here it is important to properly choose the contours in order to encircle the correct poles. This choice is dictated by the Britto-Cachazo-Feng-Witten (BCFW) recursion relations \cite{Britto:2005fq}, which of course are also based on the residue theorem. The result is that the ``top-cell'' expressions \eqref{ACCK},\eqref{MS} decompose into specific linear combinations of residues. These are themselves Yangian-invariant, and correspond to on-shell diagrams of \cite{ArkaniHamed:2012nw}. The important point now is to realize that the residue theorem is no longer available in the deformed case due to the appearance of branch cuts. So it does not make sense anymore to decompose the top-cell diagram into subsidiary on-shell components in a naive fashion, i.e.\ as though the residue theorem was still valid. Put differently, we have to give up the BCFW recursion relations, at least in the way we knew them. This is entirely consistent with the findings of \cite{Beisert:2014qba, Broedel:2014pia, Broedel:2014hca}, where it was shown that the deformed subsidiary Yangian-invariant on-shell diagrams in the non-MHV case cannot consistently be summed up to a deformed amplitude. However, this does {\it not} mean that non-MHV amplitudes cannot be deformed. It merely means that we cannot decompose them as in the undeformed case. Instead, we should take the deformed top-cell Gra{\ss}mannian integrals seriously, and consider them to yield Yangian-invariant deformations of all N$^{\hat k}$MHV tree-level amplitudes. We then have to perform the remaining integrations in the presence of branch cuts. While this certainly complicates things, there are three, related, potential benefits. Firstly, if the contours are chosen appropriately, we may hope to gain meromorphicity in the deformation parameters $\{v_j^\pm\}$, to compensate for the lost meromorphicity of the integrand on the Gra{\ss}mannian manifold. This opens up an exciting perspective: We should look for a deformed analog of the BCFW relations in the space of spectral parameters. Secondly, by way of conjecture, demanding complete analyticity of the deformed amplitude away from the poles in the $\{v_j^\pm\}$ should strongly constrain the contours. The contours of the Gra{\ss}mannian integral would be determined from a powerful principle. Thirdly, we may hope that all proper contours will be compact, and will stay away from all branch points. At loop level, this should ensure the regularization of all notorious infrared divergences, as no minors on the Gra{\ss}mannian will ever vanish along the contours.

Let us further motivate these ideas with a small mathematical Gedankenexperiment. Consider Euler's integral of the first kind, or beta function
\begin{equation}\label{beta}
B(\alpha_1,\alpha_2)=\int_{0}^1 \frac{dc}{c^{1-\alpha_1}(1-c)^{1-\alpha_2}}\,.
\end{equation}
It is well defined if ${\rm Re}\, \alpha_1>0$ and ${\rm Re}\, \alpha_2>0$. Euler showed that it equals $\Gamma(\alpha_1) \Gamma(\alpha_2)/\Gamma(\alpha_1+\alpha_2)$, where $\Gamma(\alpha)$ is his integral of the second kind, also known as the Gamma function. The result is actually a meromorphic function in both $\alpha_1$ and $\alpha_2$, a fact that is totally obscure from the integral representation \eqref{beta}.
In order to render this double-analytic continuation manifest, Pochhammer \cite{Pochhammer:1890}, not being scared by passing several times through a cut, replaced \eqref{beta} by 
\begin{equation}\label{bac}
\tilde B(\alpha_1,\alpha_2)=\frac{1}{(1-e^{2\pi i \alpha_1})(1-e^{2\pi i \alpha_2})}
\int_{\mathcal{P}} \frac{dc}{c^{1-\alpha_1}(1-c)^{1-\alpha_2}}\,,
\end{equation}
where the Pochhammer contour $\mathcal{P}$ is a closed path in the complex $c$ plane going clockwise around $c=0$, then clockwise around $c=1$, then counterclockwise around $c=0$, then counterclockwise around $c=0$, finally returning to the starting point. This continued function equals again $\Gamma(\alpha_1) \Gamma(\alpha_2)/\Gamma(\alpha_1+\alpha_2)$, but now allows for any complex values of $\alpha_1, \alpha_2 \neq \mathbb{Z}$. Poles and zeros are recovered by taking limits where the $\alpha_j$ tend to integer values. Note that the poles at which the beta function diverges have neatly factored out; the prefactor-stripped contour integral in \eqref{bac} is manifestly finite (we never come close to $c=0,1$) and manifestly analytic in the $\alpha_j$ (the contour is compact and does not care about the specific values of the $\alpha_j$).

In summary, \eqref{beta} should be a toy ``positive Gra{\ss}mannian'' integral, while \eqref{bac} should be the proper analytically continued complex version. Of course, given the integrand, meromorphicity is not sufficient. If we e.g. take a big circle around both branch points such that, for simplicity, $\alpha_1+\alpha_2=0$, we just get zero: Certainly a meromorphic function. But then we do not match the ``positive Gra{\ss}mannian'' integral. This is how positivity properties might complement meromorphicity in order to completely constrain the contours.

\section{Further Details}\label{sec:Furtherdetails}

In this section we present some details on the derivation of the deformed Gra{\ss}mannian formula \eqref{ACCK-deformed} and prove that it is invariant under the action of the level-zero and the level-one Yangian generators \eqref{deformed-yangian-symmetry}. The deformed dual Gra{\ss}mannian formula \eqref{MS-deformed} then follows through the same change of variables used in \cite{ArkaniHamed:2009vw}. As we have already pointed out, the ${\rm GL}(k)$ symmetry restricts possible deformations of
\eqref{ACCK} considerably. Let us make the following ansatz
\begin{equation}\label{ACCK-withgamma}
{\mathcal A}_{n,k}\left(\{\gamma_j\}\right)=
\int \frac{ d^{k\cdot n} C}{{\rm vol}({\rm GL}(k))}
\left(\prod_{i=1}^n(i,\,...\,,i+k-1)^{-1+\gamma_i}\right)\delta^{4k|4k}(C\cdot\mathcal{W})\,,
\end{equation}   
with $\sum_{i}\gamma_i=0$. It differs from the most general form by the fact that only cyclic minors are employed. However, we will see shortly that this suffices. Indeed, we may relate $\gamma_j$ to the evaluation representation parameters $v_j$ and central charges $c_j$ by demanding Yangian invariance of \eqref{ACCK-withgamma}. One way to proceed in order to verify this ansatz is to construct the Yangian invariants as presented in \cite{Kanning:2014maa}, see also \cite{Broedel:2014pia}. The authors of these papers generalized the approach proposed in \cite{Chicherin:2013ora}, similar to, but different from a standard Algebraic Bethe Ansatz, in order to find eigenvectors of the monodromy matrices acting on a suitable quantum space of an inhomogeneous spin chain. There is a natural classification of all such invariants by permutations $\sigma$, and we will be interested here only in the case where the invariants are associated to the shift
\begin{equation}\label{shift}
\sigma_{n,k}(i)=i+k\,\, (\mbox{mod } n).
\end{equation}
It corresponds to the aforementioned top-cell of the positive Gra{\ss}mannian ${\rm Gr}_+(k,n)$ of \cite{ArkaniHamed:2012nw}. The permutation \eqref{shift} admits the following decomposition into adjacent transpositions \cite{Broedel:2014pia}
\begin{equation}\label{decomposition}
\sigma_{n,k}=\underbrace{(k,k+1)\ldots(n-1,n)}_{}\ldots\underbrace{(23)\ldots(n-k+1,n-k+2)}\underbrace{(12)\ldots(n-k,n-k+1)}\,,
\end{equation}
where $(ij)$ denotes the transposition of the elements $i$ and $j$. Using \eqref{decomposition} one can construct Yangian invariants $|\psi\rangle_{n,k}$ for top-cells as
\begin{align}\label{Yangianinv}\nonumber
|\psi\rangle_{n,k}&=\underbrace{\mathcal{B}_{n-k,n-k+1}(y_{n-k,n-k+1})\ldots\mathcal{B}_{12}(y_{1,n-k+1})}\underbrace{\mathcal{B}_{n-k+1,n-k+2}(y_{n-k,n-k+2})\ldots\mathcal{B}_{23}(y_{1,n-k+2})}\\&\ldots\underbrace{\mathcal{B}_{n-1,n}(y_{n-k,n})\ldots\mathcal{B}_{k,k+1}(y_{1,n})}\prod_{i=1}^k \delta^{4|4}(\mathcal{W}_i),
\end{align}
where $y_{ij}=v^-_i-v^-_j$, and the $v^-_i$ are given in \eqref{v-plus-minus}. The operators $\mathcal{B}_{ij}(u)$ are formally defined in terms of complex powers $u$ of the product of super-twistor variables and their derivatives
\begin{equation}
  \label{integraloperatorB}
  \mathcal{B}_{i j}(u)=
  (-\mathcal{W}_j \cdot \partial_{\mathcal{W}_i})^u 
  =-\frac{\Gamma(u+1)}{2 \pi i}\int\frac{d\alpha}{(-\alpha)^{1+u}}
  e^{\alpha\,\mathcal{W}_j \cdot \partial_{\mathcal{W}_i}}\,,
\end{equation}
where we abbreviated $\partial_{\mathcal{W}^{\mathcal{A}}_i}\equiv \frac{\partial}{\partial{\mathcal{W}^{\mathcal{A}}_i}}$. The attentive reader should be puzzled by this complex power of a derivative operator. In fact, extensions of ordinary derivatives to operators with arbitrary powers are called fractional derivatives. They are more akin to integral operators and have manifold representations, which depend on the ranges of variables and parameters, see \cite{lavoie} for a review. We will not enter into any details here, but suggest that fractional calculus might play an important role in the construction of deformed amplitudes. Using the fact that the operators $\mathcal{B}_{ij}(u)$ act as shift operators, we may rewrite \eqref{Yangianinv} as a Gra{\ss}mannian integral and read off the powers of the minors. In a case-by-case study up to a high number of particles $n$ as well as various values for $k$, we obtained \eqref{ACCK-deformed}, up to a trivial normalization, along with the proper deformation parameters written in terms of the $v_j$ and the $c_j$ subject to the relation \eqref{cyclic-identification}. It is possible to prove \eqref{ACCK-deformed} for all $n$ and $k$ by induction, using the approach presented above. However, the proof is very technical and is omitted here. Instead, we shall simply prove Yangian invariance by directly acting with the Yangian generators on the expressions \eqref{ACCK-deformed}, \eqref{cyclic-identification} generalized from the case-by-case results.

To this purpose we will follow closely the steps of \cite{Drummond:2010qh}, appropriately adapted to our deformed case. Let us start from a Gra\ss mannian integral deformed with generic powers, see again \eqref{ACCK-withgamma}.
We notice that invariance under the level-zero generators imposes restrictions equivalent to the requirement that the measure of the Gra{\ss}mannian integral is ${\rm GL}(k)$ invariant. This leads to
\begin{equation}\label{deformed-level0}
\sum_{i=1}^n \gamma_i=0 \,.
\end{equation}
Next, let us turn to the level-one generators $\hat J$ in \eqref{deformed-yangian-symmetry} and rewrite their bilocal part as
\begin{align}\label{deformed-J1-bilocal}
 \sfrac{1}{2} \sum_{i<j} (-1)^{\mathcal C} \left[J^{\mathcal{A} \mathcal{C}}_i J^{\mathcal{C} \mathcal{B}}_j
- (i \leftrightarrow j)\right]  =  \sfrac{1}{2} \left(2 \sum_{i<j} + \sum_{i=j} - \sum_{i,j}\right)  (-1)^{\mathcal C} J^{\mathcal{A} \mathcal{C}}_i J^{\mathcal{C} \mathcal{B}}_j \,.
\end{align}
The last term is just a product of level-zero generators, and thus vanishes on the Gra\ss mannian integral. A rearrangement of the other two terms leads to
\begin{equation}\label{deformed-J1-bilocal-final}
  \sum_{i<j} \left(\mathcal{W}_i^{\mathcal{A}}\partial_{\mathcal{W}_j^{\mathcal{B}}}\mathcal{W}_j^{\mathcal{C}}\partial_{\mathcal{W}_i^{\mathcal{C}}}-\mathcal{W}_i^{\mathcal{A}}\partial_{\mathcal{W}_i^{\mathcal{B}}}\right) + \sum_{i} c_i \mathcal{W}_i^{\mathcal{A}}\partial_{\mathcal{W}_i^{\mathcal{B}}},
\end{equation}
where we again omitted level-zero generator contributions.

Along the lines of \cite{Drummond:2010qh}, the differential operators in the variables $\mathcal{W}_i^{\mathcal{A}}$  can be exchanged for  operators in the variables $c_{ai}$ when acting on the delta functions:
\begin{equation}
\mathcal{W}_j^{C}\partial_{\mathcal{W}_i^{\mathcal{C}}} \delta^{4k|4k}(C\cdot\mathcal{W}) =\left( \sum_{a=1}^k c_{ai} \frac{\partial}{\partial c_{aj}}\right) \delta^{4k|4k}(C\cdot\mathcal{W}) .
\end{equation}
The next and crucial step is to integrate by parts. Here we need to be sure that no boundary terms arise. This is ensured as long as the integration contours are closed. For open contours, one has to check that the boundary terms vanish. Proceeding under this assumption, we arrive, after some manipulations of the minors, at
\begin{align}\label{deformed-invariance-result}
\hat J^{\mathcal{A} \mathcal{B}} {\mathcal A}_{n,k}\left(\{\gamma_j\}\right) &= \sum_{b=1}^k \int \frac{ d^{k\cdot n} C}{{\rm vol}({\rm GL}(k))}
\left(\prod_{i=1}^n(i,\,...\,,i+k-1)^{-1+\gamma_i}\right) \\
& \hspace{1cm} \times \left[-\sum_{i<j}\gamma_j  + \sfrac{1}{2}\sum_{i=1}^n c_i - \sum_{i=1}^n v_i \right] \mathcal{W}_i^{\mathcal{A}}\, c_{bi}\,
\partial_{\mathcal{B}}\delta_b \prod_{m\neq b}\delta_m\,,
\end{align}
where we have defined for sake of simplicity
\begin{equation}
\delta_l := \delta^{4|4}(\sum_{i=1}^n c_{li} \mathcal{W}^{\mathcal{A}}_i) \,.
\end{equation}
Since we require this expression to vanish, we need to impose, that the term inside the square bracket be proportional to a mutual constant for every $i$
\begin{equation}
-\sum_{j=i+1}^{n}\gamma_j  + \sfrac{1}{2}  c_i  - v_i  = \beta\,, \qquad i=1,\ldots,n\,.
\end{equation}
Any such $\beta$ simply multiplies a term proportional to level-zero generators, which leads to immediate annihilation of the deformed amplitude.
This system of equations, together with \eqref{deformed-level0}, has the solution
\begin{equation}
\gamma_j=v^-_j-v^-_{j-1}, \qquad j=1,\ldots,n, \qquad    \mbox{with } \qquad v^-_{n}= -\beta \,.
\end{equation}
This is exactly the same condition we found for a large number of $n$ and $k$ by using the $\mathcal{B}$-operator method. By acting with the central charges $C_j$ on \eqref{ACCK-withgamma} we easily arrive at the relation \eqref{cyclic-identification}. This finishes the proof that \eqref{ACCK-deformed} with \eqref{cyclic-identification} is Yangian invariant.

\section{A First Look at n=6, k=3}
In this section our main focus will be on the simplest non-trivial example, namely the NMHV six-point amplitude. 
The emerging structure is already very rich and rather subtle. Here we present only a preliminary exploration, an in-depth study will be performed elsewhere.

As a warm-up exercise, let us start with the five-point NMHV amplitude, which was already successfully deformed in \cite{Ferro:2013dga} in ordinary (as opposed to momentum) twistor space. In the present context it is given by \eqref{MS-deformed} together with the integral \eqref{MS-integral}, where $n=5$ and $\hat k=1$. One immediately sees that the number of delta functions equals the number of integrations and the integral is formally evaluated by localizing it on the support of the delta functions. This yields
\begin{equation}\label{A53}
 A_{5,3}\left(\{v_j^\pm\}\right)=
\frac{\delta^{0|4}(\langle 1234\rangle \chi_5+\langle 5123\rangle \chi_4+\langle 4512\rangle \chi_3+\langle 3451\rangle \chi_2+\langle 2345\rangle \chi_1)}{\langle 1234\rangle^{1+v_1^+-v_4^-}\langle 5123\rangle^{1+v_5^+-v_3^-}\langle 4512\rangle^{1+v_4^+-v_2^-}\langle 3451\rangle^{1+v_3^+-v_1^-}\langle 2345\rangle^{1+v_2^+-v_5^-}}\,,
\end{equation}
written in terms of $4 \times 4$ determinants of four momentum-twistors
\begin{equation}\label{fourbracket}
\langle ijkl\rangle=
\epsilon_{ABCD}\,Z_{i}^{A}Z_{j}^{B}
Z_{k}^{C}Z_l^D\,,\qquad A,B,C,D=1,2,3,4\,.
\end{equation}
One observes that the result is a deformed version of the 5-cyclic so-called R-invariant
\begin{equation}
\label{R-invariant}
[ijklm]=\frac{\delta^{0|4}(\langle ijkl\rangle \chi_m+\langle jklm\rangle \chi_i+\langle klmi\rangle \chi_j+\langle lmij\rangle \chi_k+\langle mijk\rangle \chi_l)}{\langle ijkl\rangle\langle jklm\rangle\langle klmi\rangle\langle lmij\rangle\langle mijk\rangle} \,.
\end{equation}

Let us then proceed to the scattering of six particles. This corresponds to a Gra\ss mannian integral \eqref{ACCK} defined on $\mathrm{Gr}(3,6)$ in super-twistor space or, equivalently, to a $\mathrm{Gr}(1,6)$ integral in super-momentum twistor variables \eqref{MS}. In the following we will focus on the latter. It is known \cite{ArkaniHamed:2009dn} that in the undeformed case \eqref{MS} may be reduced to an integral over one variable, and that the integrand is a rational function with six poles: the amplitude is a specific combination of three residues evaluated at these poles, accomplished by choosing a suitable contour of integration. It is fixed by the BCFW recursion relation. The answer is given by a sum of three 5-cyclic terms
\begin{equation}\label{inv63}
A_{6,3}=[12345]+[12356]+[13456]\,.
\end{equation}
This result is not manifestly 6-cyclic. However, using a six-term identity, which stems from the fact that a contour enclosing all six poles yields a vanishing integral due to the rationality of the integrand, one may alternatively rewrite it in 6-cyclic form as
\begin{equation}\label{inv63cyclic}
A_{6,3}=\sfrac{1}{2}\left([12345]+[23456]+[34561]+[45612]+[56123]+[61234]\right).
\end{equation}

Let us study what happens once we introduce our deformation parameters. Since we have to abandon the BCFW recursion relations, which led to the particular combination of R-invariants in \eqref{inv63}, we do not immediately have a first-principle prescription on how to define the deformed amplitude. However, we may study the properties of the integral \eqref{MS-integral} and analyze the emergence of \eqref{inv63} as all deformation parameters tend to zero. The Gra\ss mannian integral \eqref{MS-integral} now reads
\begin{eqnarray}\label{MS-deformed-16}
 A_{6,3}\left(\{v_j^\pm\}\right) =
 \int \prod_{i=2}^6 \frac{dc_{1i}}{c_{1i}^{1-\alpha_i}}\,
\delta^{4|4}(\mathcal{Z}_1 + c_{12} \mathcal{Z}_2 +...+ c_{16} \mathcal{Z}_6)\,,
\end{eqnarray}
where we have fixed the $\mathrm{GL}(1)$ invariance by setting $c_1=1$, and put for brevity $\alpha_i = v_{i-1}^- - v_{i+1}^+$. Note again $\alpha_1 + \ldots \alpha_6=0$, which explains why the dependence on $\alpha_1$ has disappeared from \eqref{MS-deformed-16}. In order to render the integral \eqref{MS-deformed-16} well-defined we need to specify a contour of integration. As we know, \eqref{MS-deformed-16} is a formal a solution of the Yangian invariance conditions. These take the form of second order differential equations in many variables, which means that there are many linearly independent solutions. These solutions will be specified by choosing different contours. We postpone the discussion of finding appropriate contours and treat the integral formally for the moment. By saturating the four bosonic delta functions in \eqref{MS-deformed-16}, we can express any four of the variables $c_{1i}$ in terms of the remaining fifth one, which still remains to be integrated. We choose this w.l.o.g.\ to be $c_{16}$ and find the following solution
\begin{equation}
c_{1i} =  a_i + b_i \, c_{16}  \qquad i=2, ...,5\,,
\end{equation}
where $a_i$ and $b_i$ are given by ratios of momentum-twistor four-brackets \eqref{fourbracket}.
In explicit form,
\begin{align}
&a_2 = - \frac{\langle 1345\rangle}{\langle 2345\rangle}\,, \qquad a_3 = - \frac{\langle 1245\rangle}{\langle 3245\rangle}\,, \qquad a_4 = - \frac{\langle 1235\rangle}{\langle 4235\rangle} \,, \qquad a_5 = - \frac{\langle 1234\rangle}{\langle 5234\rangle}\,,\\
 &b_2 = - \frac{\langle 6345\rangle}{\langle 2345\rangle} \,,\,  \qquad b_3 = - \frac{\langle 6245\rangle}{\langle 3245\rangle} 
 \,,\, \qquad b_4 = - \frac{\langle 6235\rangle}{\langle 4235\rangle} 
 \,, \,\qquad b_5 = - \frac{\langle 6234\rangle}{\langle 5234\rangle}\,.
\end{align}
The reader may easily convince herself that, after the change of variables $c_{16} = -\sfrac{a_5}{b_5}\tau$, the remaining one-variable integral becomes
\begin{align}\label{ourlauricella}\nonumber
\mathcal{I}=&\frac{1}{\langle2345\rangle} \left(\frac{\langle 1234\rangle}{\langle 2346\rangle}\right)^{\alpha_6} \prod_{i=2}^5 a_i^{-1+\alpha_i} \int d\tau \, \tau^{-1+\alpha_6} (1-\tau)^{-1+\alpha_5} \prod_{i=2}^4  (1- z_i \tau)^{-1+\alpha_i}\\
&\times\delta^{0|4}\left( \chi_1 +\sum_{i=2}^5 (1-z_i \tau)a_i\chi_i +\frac{\langle 1234\rangle}{\langle 2346\rangle}\, \tau\,\chi_6\right),
\end{align}
with (note $z_5=1$)
\begin{equation}
z_i=\frac{a_5 b_i}{b_5 a_i}\,.
\end{equation}
The fermionic delta function is a polynomial in $\tau$ of degree four, with Gra{\ss}mann-valued coefficients. The integrand has branch points at $\tau =\infty, z_2^{-1}, z_3^{-1}, z_4^{-1}, 1,0$ for $\alpha_1, \ldots, \alpha_6 \notin \mathbb{Z}$. We notice that this integral is of hypergeometric type. It satisfies a supersymmetric version of the hypergeometric differential equation, a statement which is equivalent to the Yangian invariance of NMHV amplitudes, see section \ref{sec:Furtherdirections} below. So far we have not specified the contour, nor spelled out any possible boundaries of integration in \eqref{ourlauricella}.  As we pointed out before, this integral is Yangian invariant only if all potential boundary terms vanish when integrating by parts as in section \ref{sec:Furtherdetails}. This is trivially the case if we take a closed contour, and less trivially for open contours between any two branch points such that their associated exponents $\alpha_j$ have positive real parts. 
Note that this is not simultaneously possible for all $\alpha_j$, since their sum vanishes.
The five branch points at finite positions and the branch point at infinity divide the real line into six segments. For any two consecutive branch points $\tau_1<\tau_2$ let us define $\mathcal{I}_{(\tau_1,\tau_2)}$ to be the integral \eqref{ourlauricella} integrated between $\tau_1$ and $\tau_2$.  With a suitable change of coordinates all the allowed (i.e.\ positive real parts of the exponents at $\tau_1$ and $\tau_2$) integrals $\mathcal{I}_{(\tau_1,\tau_2)}$ may be brought to the form of the type-D Lauricella hypergeometric function, which is defined as
\begin{equation}
\label{lauricella}
F_D(\alpha,\beta_1, \beta_2, \beta_3, \gamma; z_1, z_2, z_3) = \frac{\Gamma(\gamma)}{\Gamma(\alpha) \Gamma(\gamma-\alpha)} \int_0^1 u^{\alpha-1} (1-u)^{\gamma-\alpha-1} \prod_{j=1}^3 (1-z_j u)^{-\beta_j} du \,,
\end{equation}
where convergence restricts this integral representation to ${\rm Re}\, (\alpha) >0$, ${\rm Re}\, (\gamma -\alpha)>0$.
In order to uncover some of the analytic properties of our deformed integral, let us focus on $\mathcal{I}_{(0,1)}$. 
After expanding the fermionic delta functions in \eqref{ourlauricella} and using the definition \eqref{lauricella}, we can substitute the integral with the series expansion of the type-D Lauricella hypergeometric function
\begin{equation}
F_D(\alpha,\beta_1, \beta_2, \beta_3, \gamma; z_1, z_2, z_3) =\sum_{m_1=0}^\infty \sum_{m_2=0}^\infty \sum_{m_3=0}^\infty \frac{(\alpha)_{m_1+m_2+m_3}(\beta_1)_{m_1}(\beta_2)_{m_2}(\beta_3)_{m_3}}{(\gamma)_{m_1+m_2+m_3}m_1!m_2!m_3!}z_1^{m_1}z_2^{m_2}z_3^{m_3}\,,
\end{equation}
where $(\alpha)_m$ is the (raising) Pochhammer symbol. We may now evaluate \eqref{ourlauricella} as an expansion in the $\alpha_i$ around zero. The result, up to the first subleading order, is given by
\begin{align}\label{I01}
\mathcal{I}_{(0,1)}&=\frac{1}{\alpha_6}\frac{\delta^{0|4}(\langle 1234\rangle\chi_5+\langle 5123\rangle\chi_4+\langle 4512\rangle\chi_3+\langle 3451\rangle\chi_2+\langle 2345\rangle\chi_1)}{\langle 2345\rangle^{1-\alpha_1}\langle 3451\rangle^{1-\alpha_2}\langle 4512\rangle^{1-\alpha_3}\langle 5123\rangle^{1-\alpha_4}\langle 1234\rangle^{1-\alpha_5}}\\ \nonumber
&+\frac{1}{\alpha_5}\frac{\delta^{0|4}(\langle 1234\rangle\chi_6+\langle 6123\rangle\chi_4+\langle 4612\rangle\chi_3+\langle 3461\rangle\chi_2+\langle 2346\rangle\chi_1)}{\langle 2346\rangle^{1-\alpha_1}\langle 3461\rangle^{1-\alpha_2}\langle 4612\rangle^{1-\alpha_3}\langle 6123\rangle^{1-\alpha_4}\langle 1234\rangle^{1-\alpha_6}}\\ \nonumber  
&+([12345]+[12346])\log\langle 1234\rangle-[23456]\log\frac{\langle 2346\rangle}{\langle 2345\rangle}+[13456]\log\frac{\langle 1346\rangle}{\langle 1345\rangle}\\\label{fourthline}
&-[12456]\log\frac{\langle 1246\rangle}{\langle 1245\rangle}+[12356]\log\frac{\langle 1236\rangle}{\langle 1235\rangle}+\mathcal{O}(\alpha_i) \nonumber\,,
\end{align}
where the term proportional to $\frac{1}{\alpha_6}$ is exact to all orders in  $\alpha_i$ for $i=1,\ldots,5$, and similarly for the term proportional to $\frac{1}{\alpha_5}$. We notice that the residues in front of the leading divergent terms are the deformed R-invariants as in \eqref{A53}! Clearly we could recover all possible R-invariants by focusing on other branch points. This is exciting, since we now see where the deformed lower-cell diagrams hide: They are no longer residues on the Gra{\ss}mannian manifold as in the undeformed case, but instead sit in front of poles in the space of deformation parameters. This already points towards the dissolution of the no-go theorem derived in \cite{Beisert:2014qba}. There it was shown that it is impossible to just add the deformed BCFW terms and to thereby obtain a Yangian invariant result without restricting the deformation parameters. The just derived result suggests instead, that the deformed BCFW terms should be multiplied by poles, appropriately summed, and then analytically completed by infinitely many further terms, see \eqref{I01}.

While it is clear that we have not lost any relevant information by our deformation, of course the reader would presumably still like to know how to recover the undeformed amplitude from the deformed integral in practice. Let us sketch a possible procedure. From the point of view of the differential equations given by demanding Yangian invariance, the undeformed result follows directly from setting $v_i=0$ in the definition of Yangian generators \eqref{deformed-yangian-symmetry}. Let us try to take the same limit at the level of the solutions to those equations. We need to proceed very carefully here. To demonstrate subtleties of removing the deformation, let us consider the much simpler classic hypergeometric function ${}_2 F_1$ as an example. This function gives a basis of solutions to the second order ordinary differential equation 
\begin{equation}
z(1-z)\frac{d^2 w(z)}{dz^2}+(c-(1+a+b)z)\frac{d w(z)}{dz}-a b\, w(z)=0\,.
\end{equation}
For generic values of $a,b,c$ there are two linearly independent solutions to that equation
\begin{equation}
{}_2 F_1(a,b,c,z)\qquad\mbox{and}\qquad z^{1-c} {}_2 F_1(a-c+1,b-c+1,2-c,z).
\end{equation}
However, these two solutions do not span the solution space at the ``resonant'' values of the parameters, where either of the conditions $c,c-a-b,a-b\in \mathbb{Z}$ is satisfied. In that case one has to first take a particular combination of two generic solutions, and in a subsequent step take the limit to a resonant value. We expect a similar behavior in our case -- removing the deformations corresponds to considering the resonant values of parameters. The proper combination of solutions should be given by a deformed version of the BCFW recursion relations, presumably transferred from the Gra{\ss}mannian manifold to the set of spectral planes. This will be analyzed elsewhere. 

Before closing this section let us point out that there exists a method to render the integrals $\mathcal{I}_{(\tau_1,\tau_2)}$ manifestly meromorphic in the deformation parameters $\alpha_i$ employing Pochhammer's procedure described in section \ref{meromorphicity}. We just define the analytic continuation of $\mathcal{I}_{(\tau_1,\tau_2)}$ by using Pochhammer cycles around the branch points $\tau_1$ and $\tau_2$. As a first example let us take again $\mathcal{I}_{(0,1)}$. We may then define, confer \eqref{bac},
\begin{align}\label{ourlauricella_ac}\nonumber
\mathcal{\tilde I}_{(0,1)}=&\frac{1}{(1-e^{2\pi i\alpha_6})(1-e^{2\pi i \alpha_5})}\frac{1}{\langle2345\rangle} \left(\frac{\langle 1234\rangle}{\langle 2346\rangle}\right)^{\alpha_6} \prod_{i=2}^5 a_i^{-1+\alpha_i} \int_{\mathcal{P}_{(0,1)}} d\tau \, \tau^{-1+\alpha_6} (1-\tau)^{-1+\alpha_5} \\
&\times\prod_{i=2}^4  (1- z_i \tau)^{-1+\alpha_i}\,\delta^{0|4}\left( \chi_1 +\sum_{i=2}^5 (1-z_i \tau)a_i\chi_i +\frac{\langle 1234\rangle}{\langle 2346\rangle}\, \tau\,\chi_6\right),
\end{align}
where $\mathcal{P}_{(0,1)}$ is the Pochhammer contour snaking around the branch points at $0$ and $1$.
This integral agrees with $\mathcal{I}_{(0,1)}$ as long as $\mathrm{Re} \, \alpha_5 > 0$ and $\mathrm{Re} \, \alpha_6 > 0$. The question we have not fully analyzed yet is how to reassemble these building blocks into the ``correct'' multi-meromorphic function corresponding to the properly deformed amplitude. Here a matching to the ``positive Gra{\ss}mannian'' with ``positive'' external data is presumably sufficiently constraining.

\section{Further Directions}
\label{sec:Furtherdirections}

In the previous section we have encountered the deformation of the $\mathcal{A}_{6,3}$ amplitude in terms of Lauricella hypergeometric functions. It turns out that there is a broader class of hypergeometric functions, introduced by Gelfand \cite{MR841131}, which are very closely connected to our deformations\footnote{The relevance of Gelfand hypergeometric functions, as described in \cite{Aomoto,vilenkin1993representation},
as well as the relation to the qKZ equations were independently noticed by Nils Kanning and Rouven Frassek.}.
In this section we will sketch possible relations between the two. General hypergeometric functions also make their appearance as solutions to the Knizhnik-Zamolodchikov equation. We suggest how the latter may be related to Yangian invariants. 

First of all, let us emphasize that Yangian invariants are solutions to a particular set of differential equations of first and second order. In the case of NMHV amplitudes\footnote{We suspect that there exists an even larger class of hypergeometric differential equations satisfied by the general ${\rm N}^{k}{\rm MHV}$ amplitudes. However, we were not able to find these in the mathematical literature.} written in momentum twistor space these equations may be elegantly rewritten as
\begin{equation}\label{hyperseteq}
\begin{cases}
& \sum_{\mathcal{A}}^{} \mathcal{Z}^\mathcal{A}_j \frac{\partial}{\partial  \mathcal{Z}^\mathcal{A}_j} F = \alpha_j F\,, \\
& \sum_{j} \mathcal{Z}^\mathcal{A}_j \frac{\partial}{\partial  \mathcal{Z}^\mathcal{B}_j} F = -(-1)^{\mathcal{A}} \delta_{\mathcal{A}\mathcal{B}} F\,, \\
& \frac{\partial^2}{\partial\mathcal{Z}^\mathcal{A}_j \partial\mathcal{Z}^\mathcal{B}_i} F =  \frac{\partial^2}{\partial\mathcal{Z}^\mathcal{A}_i \partial\mathcal{Z}^\mathcal{B}_j} F\,,
\end{cases}
\end{equation}
where $F$ is the $\mathrm{Gr}(1,n)$ Gra\ss mannian integral \eqref{MS-integral}. 
The first set of equations is the statement of homogeneity of $F$ in the $\mathcal{Z}_j^\mathcal{A}$ variables, where the $\alpha_j$ are related to the representation labels $c_j$ and $v_j$. The second group of equations is the statement of $\mathfrak{gl}(4|4)$ (or more generally $\mathfrak{gl}(N|M)$) invariance of $F$. The third set may be interpreted as the action of the level-one Yangian generators when written in momentum twistor space.
A similar set of equations arises for the bosonic algebras $\mathfrak{gl}(N)$ in the definition of the Gelfand hypergeometric functions, see \cite{Aomoto,vilenkin1993representation} for introductions to this subject. These are hypergeometric functions in several variables. They possess representations in terms of complex integrals of complex powers of polynomials, and are naturally associated to Gra\ss mannians. Let us note that N$^{\hat k}$MHV amplitudes for $\hat k>1$ do not satisfy  \eqref{hyperseteq}. It would be intriguing to find a more general version of these differential equations allowing for arbitrary $\hat k$.

Another interesting observation is a link between Yangian invariants and the solutions to the quantum version of the Knizhnik-Zamolodchikov (qKZ) equation \cite{frenkel1992}, which appears e.g.~as a constraint on correlation functions of vertex operators in two-dimensional integrable conformal field theories. Let us consider a function $\Phi(z_1,\ldots,z_n)$ with values in a tensor product $V_1\otimes \ldots\otimes V_n$ of highest weight $\mathfrak{gl}(N|M)$-modules. The qKZ equation is a system of difference equations satisfied by $\Phi$ of the form
\begin{equation}\label{qKZ}
\Phi(z_1,\ldots,z_i+p,\ldots,z_n)=K_i(z_1,\ldots,z_n;p)\, \Phi(z_1,\ldots,z_n)
\end{equation}
with the qKZ operators $K_i$ given by
\begin{equation}
K_i(z_1,\ldots,z_n;p)=L_{i\,i-1}(z_i-z_{i-1}+p)\ldots L_{i\,1}(z_i-z_1+p)L^{-1}_{n\,i}(z_n-z_i)\ldots L^{-1}_{i+1\,i}(z_{i+1}-z_i)\,.
\end{equation}
The operators $L_{ij}(z)$ are intertwiners corresponding to pairs $V_i,V_j$ and $p\in\mathbb{C}$. The solutions to the system of equations \eqref{qKZ} can be found using the Algebraic Bethe Ansatz technique \cite{Reshetikhin1992}, since they are related to the eigenvectors of a suitable transfer matrix defined on an inhomogeneous spin chain. In \cite{Frassek:2013xza} it was shown that also the Yangian invariance condition can be rewritten as an eigenproblem for such a spin chain, where the Yangian invariants are the eigenvectors of the monodromy matrix. Since the transfer matrix may be obtained from the monodromy matrix by taking the trace over an auxiliary vector space, this should result in a relation between Yangian invariants and the solutions $\Phi$ in \eqref{qKZ}. It would be very interesting to make this relation explicit.  

Interestingly, the classical limit of the qKZ equation, the ordinary Knizhnik-Zamolodchikov (KZ) equation \cite{Knizhnik198483}, appeared already in the context of scattering amplitudes in $\mathcal{N}=4$ SYM in the direct Feynman diagram calculations \cite{Henn:2013tua}. This is not surprising, since it is closely related to polylogarithm functions, which form a functional basis for loop-level results in any quantum field theory. It should be instructive to further investigate this relation.

One may suspect that the deformation of the ${\cal N}=4$ Yangian-invariant Gra{\ss}mannian contour integrals will also work in the case of the planar three-dimensional ${\cal N}=6$ super-conformal Chern-Simons model: A Yangian-invariant Gra{\ss}mannian integral formula for this so-called planar ABJM theory was derived in \cite{Lee:2010du}, and much of the on-shell diagram formalism of \cite{ArkaniHamed:2012nw} carries over from the four- to the three-dimensional model. This is indeed the case, as was very recently shown in \cite{Bargheer:2014mxa}. The authors also report an independent derivation of our ${\cal N}=4$ expressions \eqref{ACCK-deformed} and \eqref{MS-deformed},\eqref{MS-integral}.

\section{Outlook}

Clearly, the integrable deformation of the Gra{\ss}mannian approach to scattering amplitudes is of great mathematical interest. It is fairly obvious that the deformed integrals lead to generalized multi-variate hypergeometric functions. This is a rich subject intensively investigated by mathematicians from the mid-18th century all the way to the present. From the physics perspective, we need to establish that the deformed Gra{\ss}mannian integrals are useful for loop calculations. We hope that they will lead to a deepened analytic understanding of all radiative corrections to the tree-level amplitudes, while staying in exactly $1+3$ dimensions. We feel very encouraged by the interesting work of Penrose and Hodges, who considered the very same deformations already since the early days of the twistor approach, with the goal of regulating massless scattering processes in quantum field theory \cite{Penrose:1972ia,Hodges:1985ac}. The only missing elements were supersymmetry and integrability. In this context the next step, apart from in-depth investigations of various special cases, might be to directly deform the amplituhedron of \cite{Arkani-Hamed:2013jha, Arkani-Hamed:2013kca}.

\section*{Acknowledgments}
We thank Nima Arkani-Hamed, Johannes Brödel, James Drummond, Rouven Frassek, Andrew Hodges, Nils Kanning, Yumi Ko, Arthur Lipstein, David Meidinger, Gregor Richter, and Jaroslav Trnka for useful discussions. 
This research is supported in part by the SFB 647 \emph{``Raum-Zeit-Materie. Analytische und Geometrische Strukturen''} and the Marie Curie network GATIS (\texttt{\href{http://gatis.desy.eu}{gatis.desy.eu}}) of the European Union’s Seventh Framework Programme FP7/2007-2013/ under REA Grant Agreement No 317089. T.L.\ is supported by ERC STG grant 306260. L.F. \ is supported by the Elitenetwork of Bavaria. M.S.\ thanks the Theory Group at CERN for hospitality during a precious sabbatical. 


\bibliographystyle{nb}
\bibliography{bibliography}
\end{document}